\documentclass{article}

\usepackage{arxiv}
\usepackage[utf8]{inputenc} 
\usepackage[T1]{fontenc}    
\usepackage{url}            
\usepackage{booktabs}       
\usepackage{amsfonts}       
\usepackage{nicefrac}       
\usepackage{microtype}      
\usepackage{titling}
\usepackage{amsmath}
\usepackage{graphicx}

\title{{\normalsize A Technical Report on Preliminary Investigations of}\\
       {Linear Stability Analysis of a Normal Shock Train in a Constant Area Isolator of a Hypersonic ScramJet}}

\author{
    {Swetava Ganguli\thanks{Department of Computer Science, Stanford University, Stanford, CA 94305, and Department of Mechanical Engineering, Stanford University, Stanford, CA 94305}} \\
    {\texttt{swetava@cs.stanford.edu}}
}

\date{}

\begin{document}

\maketitle

\begin{abstract}
Recent experiments in hypersonic vehicle technology have had some successful outcomes (ex. NASA X-43, Boeing X-51) but also have seen some unexpected failures (DARPA Hypersonic Test Vehicle 2). One important failure mode is the problem of engine unstart. \cite{Morgan2012} has discussed and analyzed a Scramjet isolator using two unit problems, viz. that of an oblique shock impinging on a turbulent boundary layer and that of a normal shock train in a constant-area duct. In order to analyze unstart, we use the numerical data obtained from the LES calculations of the normal shock train in a constant area duct and perform temporal linear stability analysis on it under the assumptions of inviscid semi-parallel flow. This technical report describes the methodology and the details of this linear stability analysis. The results obtained from the investigations have been documented and interpreted with their relevance to the physical problem. The report concludes that temporal linear stability analysis is insufficient to analyze the problem of unstart. Finally, the need for a new approach (namely, dynamic mode decomposition (DMD)) has been discussed.
\end{abstract}

\section{Introduction}
Supersonic combustion in ramjets or scramjets may potentially prove to be more efficient than conventional ramjet engines at speeds higher than Mach 7.0. As described by \cite{Morgan2012}, the basic concept of a scramjet is as follows: first air is captured through an inlet and decelerated through the action of oblique shock waves to lower, but still supersonic Mach numbers. An isolator section may be utilized to allow the flow to adjust to a static back pressure higher than the static inlet pressure. Combustion then takes place in the stabilized supersonic stream, forming a pre-combustion shock train in the isolator. The after-combustion flow is then expanded through a nozzle, and thrust is obtained.

Historically, there has been a lot of interest in hypersonic propulsion. However, high-fidelity computational science and powerful supercomputers have refueled some of that interest. Successful flight experiments such as the NASA X-43, Boeing X-51, and University of Queensland HyShot programs have recently demonstrated the feasibility of scramjet propulsion at flight conditions up to about Mach 10. However, the recent (earlier part of this decade), unexpected failures of the DARPA Hypersonic Test Vehicle 2, as well as earlier failures in the aforementioned programs have underscored the need for improved predictive capability in simulation. As described in \cite{Morgan2012}, one particular failure mode that has frustrated many programs is the phenomenon of engine unstart. One avenue of research towards a better predictive capability of unstart design margins is investigation of the physics of shock/turbulence interaction in the isolator.

\textit{Unstart} is the phenomenon in which the isolator shock system is ejected through the inlet. Unstart occurs when significant regions of subsonic flow develop in the combustor and propagate upstream through the isolator. Although unstart is often triggered by an over-injection of fuel or a change in fuel mixing properties (a phenomenon often called thermal choking), it can also be induced by an increase in back-pressure or by excessive boundary layer separation in the isolator. This emphasizes that one avenue of research towards a better predictive capability of unstart design margins is investigation into the physics of shock/turbulence interaction in the isolator. \cite{Morgan2012} explores, using high-fidelity LES, two unit problems of shock/turbulence interaction often found in scramjet isolator systems: (1) that of an oblique shock impinging on a turbulent boundary layer, and (2) that of a normal shock train in a constant-area duct.

It is anticipated that unstart may occur due to disturbances propagating upstream due to instabilities of the shock train. The aim of the present research is to investigate the stability of the normal shock train in a constant-area duct using the LES data from calculations in \cite{Morgan2012} as the mean flow. We aim to understand the nature of the disturbances (if they are present) arising due to these instabilities. The first step is a simplified approach, that of linear stability analysis assuming parallel flow in the duct in the inviscid limit. The present report outlines the details of the linear stability analysis, the methodology used to solve the resulting eigenvalue problem, the nature of the instabilities that result from this analysis, and conclusions that we can draw based on this analysis.

\section{Temporal Stability Analysis}

\subsection{Governing Equations}
We aim to perform temporal stability analysis on the normal shock train in a constant-area duct. We will use the flow profiles (time-averaged) arising from LES calculations in \cite{Morgan2012} as the mean flow for the stability analysis. This is done by choosing representative cross-sections of the shock train regions. Note that the LES calculations are three-dimensional and we choose the planar depth-wise cross-section as our two-dimensional base flow. Care has been taken to choose these sections some distance from the shocks so that the strong viscous effects considerably die out and the inviscid assumption holds in at least a weak sense. Furthermore, for the first few shocks in the shock train region, the ratio of the transverse velocity ($v$) to the stream-wise velocity ($u$) barely exceeds 3\%. Thus, a parallel flow assumption is justified in this region. However, as we move downstream, this assumption breaks down. After completing our analysis based on the parallel flow assumption, we will analyze the impact of this assumption on the results of the analysis.     

The working fluid is non-dissociating, non-reacting air. Therefore, we consider the equations of motion of an ideal Newtonian fluid, governed by the conservation of mass, momentum, and energy.  Although these equations are given in a variety of texts they are reproduced here for completeness and to introduce the notation that will be used throughout the remainder of this study. 

In two-dimensional Cartesian coordinates, we denote position and velocity vectors by $x = (x, y)$ and $u = (u, v)$, respectively where the boldface denotes a vector and the components are shown in the brackets. Then, the conservation of mass may be expressed by the continuity equation, given by equation (\ref{equation_1}), conservation of momentum is given by the Navier-Stokes equations, reproduced in equations (\ref{equation_2}) for the x-direction and (\ref{equation_3}) for the y-direction, and conservation of energy is given in equation (\ref{equation_4}). Non-dimensionalization of space, density, velocity, pressure and temperature is carried out with respect to dimensional reference conditions $L$, $\rho_0$, $c_0 = \sqrt{\gamma R T_0}$, $\rho_0 c_0^2$, and $T_0$ respectively. Thus the reference velocity is the speed of sound at the reference conditions $\rho_0$, $T_0$.

\begin{equation}\label{equation_1}
    \frac{\partial \rho}{\partial t} + \frac{\partial \rho u}{\partial x} + \frac{\partial \rho v}{\partial y} = 0
\end{equation}

\begin{equation}\label{equation_2}
    \frac{\partial \rho u}{\partial t} + \frac{\partial \rho u u}{\partial x} + \frac{\partial \rho u v}{\partial y} = -\frac{\partial p}{\partial x}
\end{equation}

\begin{equation}\label{equation_3}
    \frac{\partial \rho v}{\partial t} + \frac{\partial \rho u v}{\partial x} + \frac{\partial \rho v v}{\partial y} = -\frac{\partial p}{\partial y}
\end{equation}

\begin{equation}\label{equation_4}
    \frac{\partial \rho E}{\partial t} + \frac{\partial \rho u E}{\partial x} + \frac{\partial \rho v E}{\partial y} = \frac{\partial p}{\partial t}
\end{equation}

\begin{equation}\label{equation_5}
    \gamma p = \rho T 
\end{equation}

The fluid, air, is assumed to obey the ideal gas equation of state given by equation (\ref{equation_5}). The total energy is defined as given in equation (\ref{equation_6}). The ratio of specific heats for air is given by $\gamma = 1.4$.

\begin{equation}\label{equation_6}
    \rho E = \frac{\rho^* E^*}{\rho_0 c_0^2} = \frac{\rho^* \left(e^* + k^* + p^*/\rho^*\right)}{\rho_0 c_0^2} 
\end{equation}

\subsection{Linearization and Linearized Perturbation Equations}
In order to linearize the non-linear equations governing the flow, we introduce small disturbances (small compared to mean flow) to the mean flow and study the evolution of these disturbances in time (temporal stability). This process may be seen as resolving the total flow into an averaged mean flow and a fluctuating component \cite{Criminale67,Criminale03}. In the calculation that follows, the non-linear terms are generated from products of fluctuating velocities and their derivatives. These interactions will only modify the non-fluctuating flow or introduce higher harmonics. Hence, in the linearization process we neglect: 
\begin{enumerate}
    \item Second or higher order multiplication terms
    \item Products of fluctuations/perturbations
    \item Derivatives of fluctuations/perturbations
\end{enumerate}
We denote by the tilde superscript the fluctuating components of the flow. Then the total flow properties may be written in terms of the mean/base flow and the fluctuating components as

\begin{equation}\label{equation_7}
    \begin{split}
        u    &= U + \tilde{u}       \\
        v    &= \tilde{v}           \\
        p    &= P + \tilde{p}       \\
        T    &= T + \tilde{T}       \\
        \rho &= \rho + \tilde{\rho}
    \end{split}
\end{equation}

Here, $U$, $P$, $T$, and $\rho$ are the mean/base stream-wise velocity, pressure, temperature, and density fields. Note that in accordance with the parallel flow assumption, the mean transverse velocity field is indeed zero.  On substituting (\ref{equation_7}) into equations (\ref{equation_1}) through (\ref{equation_6}), we obtain

\begin{equation}\label{equation_8}
    \frac{\partial \tilde{\rho}}{\partial t} + \frac{\partial \left(\rho \tilde{u} + \tilde{\rho} u\right)}{\partial x} + \frac{\partial \rho \tilde{v}}{\partial y} = 0
\end{equation}

\begin{equation}\label{equation_9}
    \rho \frac{\partial \tilde{u}}{\partial t} + \rho u\frac{\partial  \tilde{u}}{\partial x} + \rho \tilde{v}\frac{\partial u}{\partial y}  + \frac{\partial \tilde{p}}{\partial x} = 0
\end{equation}

\begin{equation}\label{equation_10}
    \rho\frac{\partial \tilde{v}}{\partial t} + \rho u\frac{\partial \tilde{v}}{\partial x} + \frac{\partial \tilde{p}}{\partial y} = 0
\end{equation}

\begin{equation}\label{equation_11}
    \frac{1}{\left(\gamma - 1\right)}\left\{\frac{\partial \rho \tilde{T}}{\partial t} + \frac{\partial \rho u \tilde{T}}{\partial x} + \rho \tilde{v}\frac{\partial T}{\partial y}\right\} = \frac{\partial \tilde{p}}{\partial t} + u\frac{\partial \tilde{p}}{\partial x}
\end{equation}

\begin{equation}\label{equation_12}
    \gamma \tilde{p} = \rho \tilde{T} + \tilde{\rho} T
\end{equation}

We will now take recourse to the method of eigenfunction expansions in terms of the Fourier modes of the disturbances. The linearity of the partial differential equations (PDEs) in equations (\ref{equation_8}) through (\ref{equation_11}) gives us this privilege.

\subsection{Method of Eigenfunction Expansion or Normal Mode Analysis}
We shall immediately take advantage of linearity and seek solutions in terms of complex functions. This will allow us to reduce these PDEs into ordinary differential equations that lend themselves to simpler analysis. We assume that the perturbations have the form of a travelling wave with wavenumber component $\alpha$ in the x direction and frequency/growth rate $\omega = \alpha c$ in time. Thus, we hope to find solutions of the form (ansatz)

\begin{equation}\label{equation_13}
    \begin{split}
        \tilde{u}    &= f(y)\exp\left(i \alpha \left(x - ct\right)\right)           \\
        \tilde{v}    &= \alpha \phi(y)\exp\left(i \alpha \left(x - ct\right)\right) \\
        \tilde{p}    &= \pi(y)\exp\left(i \alpha \left(x - ct\right)\right)         \\
        \tilde{T}    &= \theta(y)\exp\left(i \alpha \left(x - ct\right)\right)      \\
        \tilde{\rho} &= r(y)\exp\left(i \alpha \left(x - ct\right)\right) 
    \end{split}
\end{equation}

In general, $f, \phi, \pi, \theta, r, \alpha$, and $c$ are complex quantities. The actual solution is the real part of the addition of the quantities in (\ref{equation_13}) which is nothing but half the sum of the quantities in (\ref{equation_13}) and their complex conjugates. It suffices therefore to investigate the nature of the perturbations with only the forms shown in (\ref{equation_13}) keeping in mind that the actual properties that we seek are the real components of these quantities.

In temporal stability analysis, the disturbances are assumed to amplify in time and not space. Hence, $c$ (the complex phase speed = $c_r + ic_i$) becomes the unknown eigenvalue, and $\alpha$, the wavenumber (spatial frequency) is specified as a real number. Thus, we have temporal instability if $c_i > 0$, neutral stability if $c_i =0,$ and temporal stability if $c_i < 0$. The above eigenfunction expansion transforms the problem into an eigenvalue problem which may be expressed as

\begin{equation}\label{equation_14}
    c = F\left(\alpha, M = U_{\text{mean}}/c_0\right)
\end{equation}

where $F$ is a complex map. The above equation yields a pair $c = (c_r, c_i)$ when $\alpha$ and $M$ are specified. The phase speed is simply $c_r$  and $M = 1$ (based on the chosen non-dimensionalization as described in section 2.1) in this case. The ordinary differential equations (ODEs) that result from the substitution of (\ref{equation_13}) into (\ref{equation_8}) through (\ref{equation_10}) are 

\begin{equation}\label{equation_15}
    \begin{gathered}
        i(U-c)r + i\rho f + \left(\rho \phi\right)' = 0                    \\
        \rho\left[i(U-c)f + U'\phi\right] + i\pi =0                        \\
        i\alpha^2\rho(U-c)\phi + \pi' = 0                                  \\
        \rho\left[i(U-c)\theta + T'\phi\right] - i(\gamma - 1)(U-c)\pi = 0 \\
        \gamma\pi = \rho\theta + rT 
    \end{gathered}
\end{equation}

which may be further reduced to two first order equations 

\begin{equation}\label{equation_16}
    \begin{gathered}
        \alpha^2(U-c)(\rho\phi T) = i\pi'T                      \\
        (\rho\phi T)'(U-c) - U'(\rho\phi T) = i\pi(T - (U-c)^2)
    \end{gathered}
\end{equation}

which may then be written as a second order ODE for the pressure disturbance as

\begin{equation}\label{equation_17}
    \frac{d^2 \pi}{dy^2} + \left[\frac{1}{T}\frac{dT}{dy} - \frac{2}{U-c}\frac{dU}{dy}\right]\frac{d\pi}{dy} - \frac{\alpha^2}{T}\left(T - (U-c)^2\right)\pi = 0
\end{equation}

We recognize that this ODE is a Ricatti-type ODE and use the Ricatti transformation to transform the second order ODE into a first order boundary value problem as 

\begin{equation}\label{equation_18}
    \frac{dG}{dy} = -\alpha T G^2 - \left[\frac{2}{T}\frac{dT}{dy} - \frac{2}{U-c}\frac{dU}{dy}\right]G + \frac{\alpha}{T^2}\left(T - (U-c)^2\right)
\end{equation}

where 

\begin{equation*}
    G = \frac{1}{\alpha \pi T}\frac{d\pi}{dy}
\end{equation*}

is the Ricatti Transformation Variable (RTV). In order to completely state the problem, we must impose boundary conditions on the boundary value problem.  That will be the focus of the next subsection.

\subsection{Boundary Conditions}
\begin{figure}
    \centering
    \includegraphics[width=0.7\textwidth]{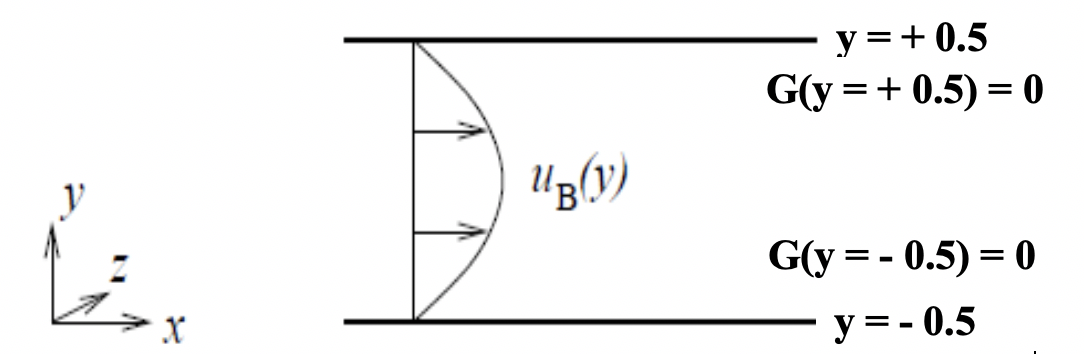}
    \caption{Description of problem geometry and mean flow conditions. Also shown are the boundary conditions. \label{figure_1}}
\end{figure}

Refering to figure \ref{figure_1}, at $y = -0.5$ and $y = +0.5$, the velocities and its perturbations are zero on the walls of the duct due to the \textit{no-slip} boundary condition. From the first order equations governing the flow, (\ref{equation_16}), it may concluded that $\phi = 0 \implies \pi' = 0$. Here, since all derivatives are with respect to $y$, the prime denotes a derivative with respect to $y$. Thus we have

\begin{equation}\label{equation_19}
    \begin{gathered}
        G(y = +0.5) = 0 \\
        G(y = -0.5) = 0
    \end{gathered}
\end{equation}

\subsection{Complete Eigenvalue Problem}
If the mean flow profiles are given, as has been from the LES data, then the boundary conditions (\ref{equation_19}) and the Ricatti equation (\ref{equation_18}) completely specify our boundary value problem. In addition, on choosing a real $\alpha$ (as is required for temporal stability analysis), we have our eigenvalue problem for $c$.

\section{Methodology and Implementation}

\subsection{Algorithm to solve Ricatti eigenvalue problem}
The extra effort to convert the Rayleigh equation for the pressure disturbance (\ref{equation_17}) to a Ricatti equation (\ref{equation_18}) offers several advantages. Most notable is the fact that Rayleigh's equation is homogeneous and so has the trivial solution in its solution space. Unless one is close to the correct eigenvalue, the solution may converge only to the trivial solution. The Riccati equation is non-homogeneous and hence does not have the trivial solution in its solution space.

The numerical procedure adapted for solving for the eigenvalues is as follows \cite{Criminale03}:
\begin{itemize}
    \item[Step 1:] Fix a value of the wavenumber $\alpha$ (real) and make an initial estimate on the phase speed (complex) $c$
    \item[Step 2:] Integrate the Ricatti Equation from $-L/2$ to 0 and from $L/2$ to 0. Note that we have the boundary conditions specified at $-L/2$ and $L/2$. For our case of the channel with the choice of the dimensionless parameters, $L = 1$. We will use the Runge-Kutta method of fourth order (RK4) with adaptive stepping to carry out this integration.
    \item[Step 3:] Compute $\left[G\right]\bigg|_{y=0} = G(y=0^+) - G(y=0^-)$, the jump in $G$ at the origin
    \item[Step 4:] If $[G] < \text{tol}$, where tol is a small number, then stop. The value of $c$ is the required eigenvalue. If $[G] > \text{tol}$, compute an updated value of $c$ by some numerical root finding procedure. Go to Step 2 and repeat until convergence
\end{itemize}

We use the MATLAB optimization toolbox to seek the eigenvalue and drive the jump to an interval near zero whose length is determined by the tolerance \textit{tol}. The tolerance specified for the jump in our implementation is $1$ x $10^{-12}$ and that for the eigenvalue ($c$) is $1$ x $10^{-8}$. Our description of the analytical work is now complete. It is now only left to describe the procedure by which the base flow has been obtained from the LES data from \cite{Morgan2012}.

\subsection{Extraction of Mean Flow from LES Calculations in [1]}
\begin{figure}[ht]
    \centering
    \includegraphics[width=\textwidth]{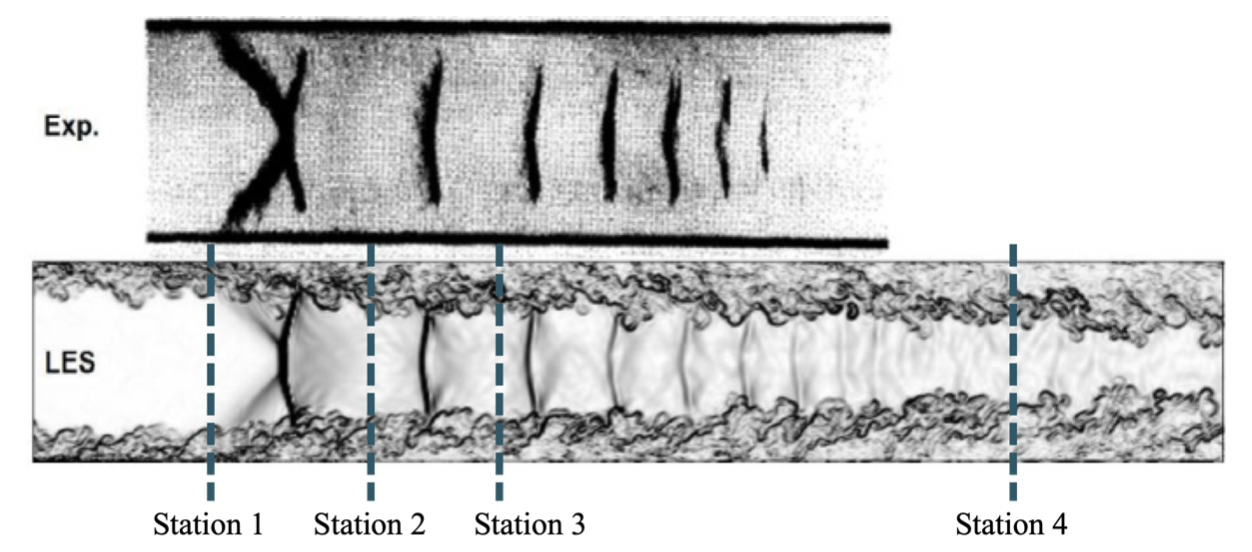}
    \caption{(Top) Experimental Schlieren image of a shock train at $M = 1.61$, $Re_\theta = 16200$ \cite{Carroll88}. (Bottom) Numerical Schlieren (instantaneous magnitude of density gradient) from LES with spanwise periodic boundary conditions at $M = 1.61$, $Re_\theta = 1660$ \cite{Duraisamy12}. Also shown are the stations used for the temporal stability analysis.  \label{figure_2}}
\end{figure}

For stability analysis, the velocity, pressure and density profiles have been extracted at 8 stations (refer figure \ref{figure_2}) corresponding to positions upstream and downstream of the individual shocks in the shock train. Temperature is then calculated from the ideal gas equation of state. Four of those stations have been shown in figure 2. Temporal stability analysis has been performed at these 8 stations individually. This choice of 8 representative stations shows us the trend of the stability as we progress downstream of the first shock (strongest shock).  Furthermore, we can investigate the changes that occur inside the shock train region based on spatial distance and perhaps comment on the `most' unstable and the `most' stable regions inside the shock train. If such a map can be made, appropriate measures may later be devised that may selectively damp out these disturbances. If indeed unstart is a result of these disturbances, we would have a quantitative method of eliminating this problem.  

\section{Results}
In all the computations, we have varied the wavenumber from 0.1 to 1 at an interval of 0.01. Variations of relevant quantities have been plotted and shown together as a comparison between the stations.

\subsection{The variation of phase speed with wavenumber}
\begin{figure}
    \centering
    \includegraphics[width=\textwidth]{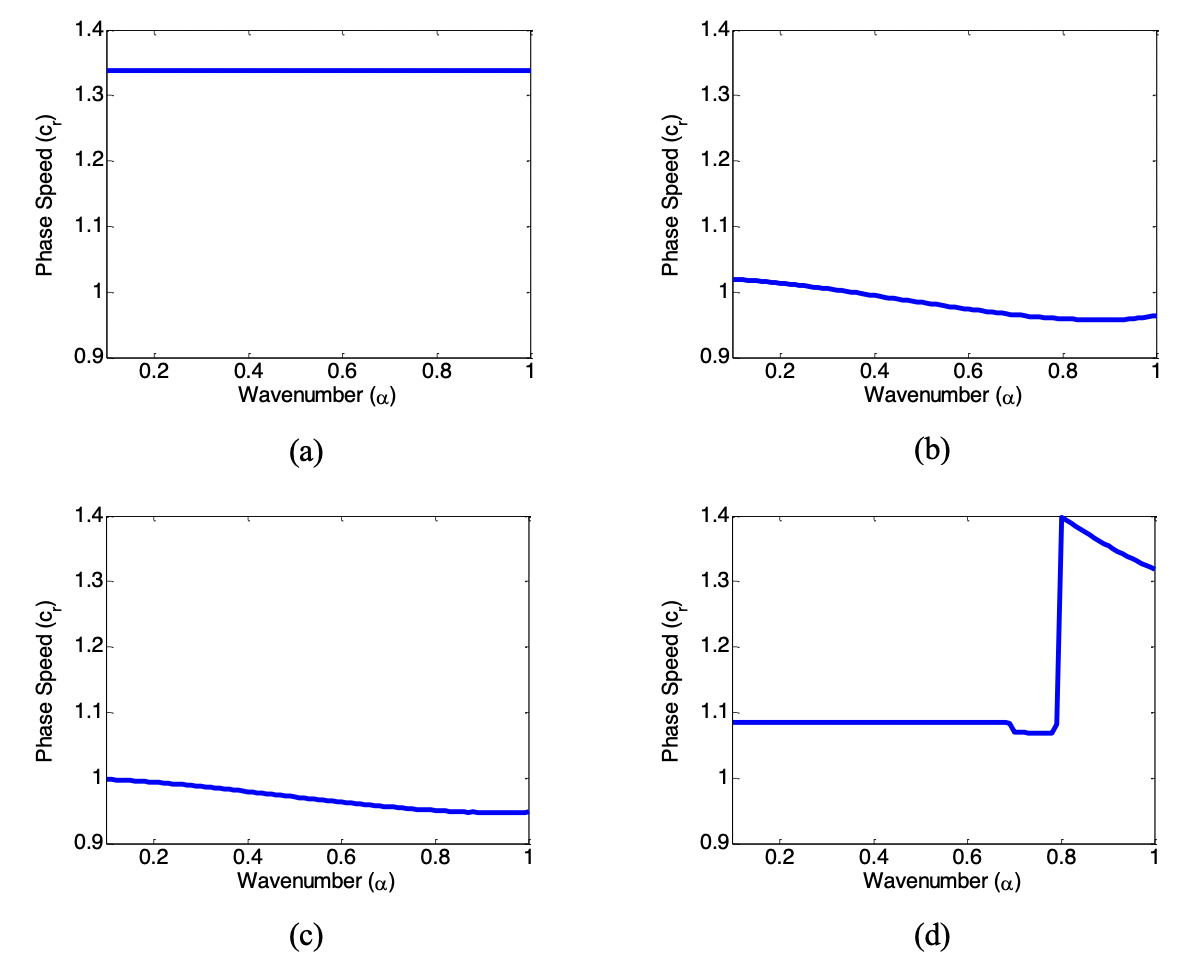}
    \caption{The variation of phase speed with respect to wavenumber ($\alpha$) has been shown for (a) Station 1, (b) Station 2, (c) Station 3, and (d) Station 4.  \label{figure_3}}
\end{figure}

It can be seen from figure \ref{figure_3} that all the phase speeds are positive. This means that all disturbances advect downstream along with the mean flow. There are no upstream propagating disturbances that are linearly unstable. 

\subsection{The variation of growth rate with wavenumber}
Station 1, which is a station upstream of the first shock, is seen in figure 4 to have a negative growth rate throughout in the inspected wavenumber region. Furthermore, the trend is indeed towards more negative values. Thus, this station is linearly stable for all wavenumbers. This story however changes as we proceed further downstream to the next station. Here, we can see an increasing growth rate between certain range of wavenumbers after which the growth rate peaks. This maximum growth rate denotes the fastest growing disturbance and the corresponding wavenumber identifies this most unstable disturbance. Beyond this peak, we have a trend of decreasing growth rates. As we proceed further downstream, the same qualitative behavior of the growth rate with respect to wavenumber is seen. However, the absolute value of the growth rate decreases at all wavenumbers and is almost zero at station 4, which is the station downstream of the shock train region. Thus, it may be concluded that the region just downstream of the first (strongest) shock is the most unstable region.    

\begin{figure}
    \centering
    \includegraphics[width=\textwidth]{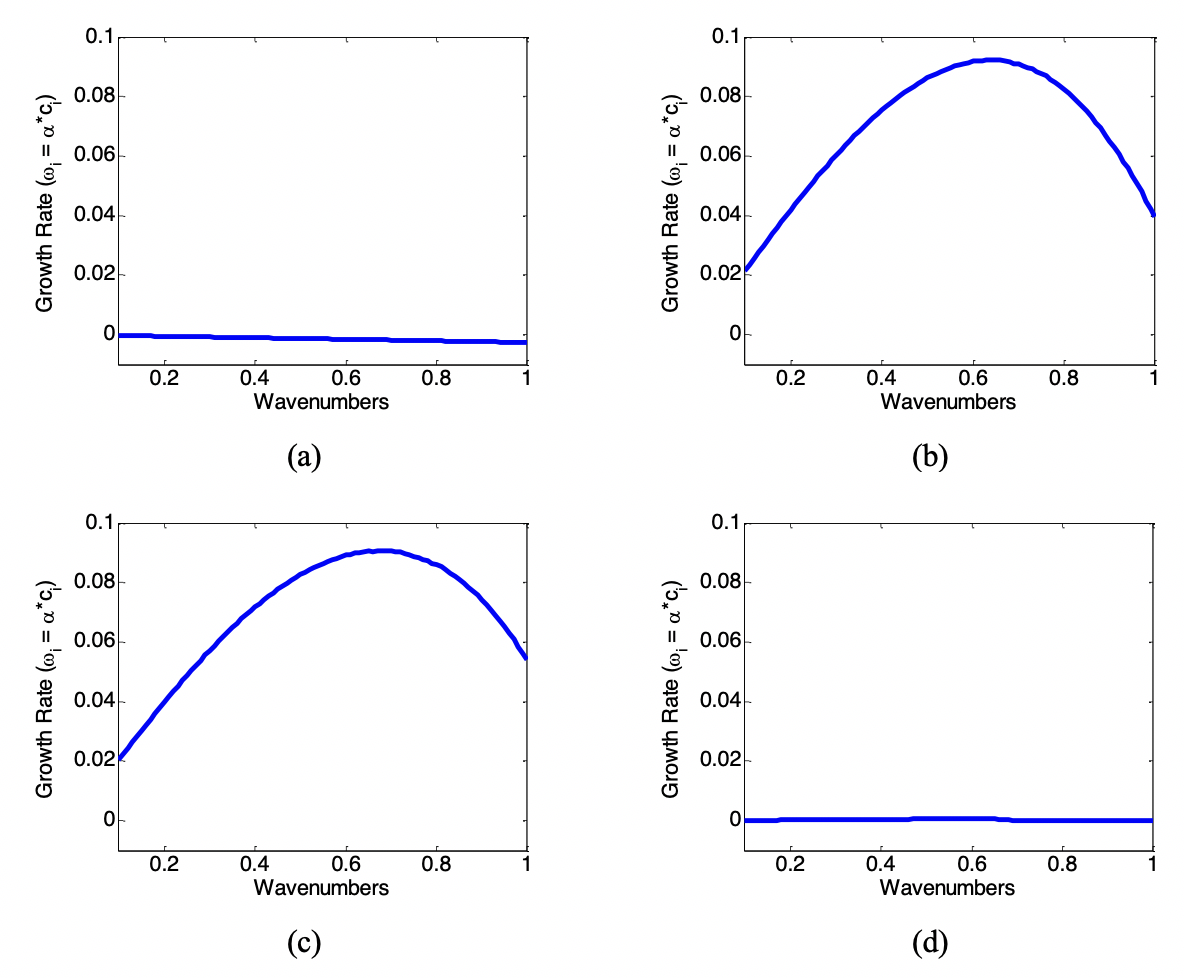}
    \caption{The variation of growth rate with respect to wavenumber has been shown for (a) Station 1, (b) Station 2, (c) Station 3, and (d) Station 4.  \label{figure_4}}
\end{figure}

\subsection{The variation of the pressure disturbance with wavenumber and space}
\begin{figure}
    \centering
    \includegraphics[width=\textwidth]{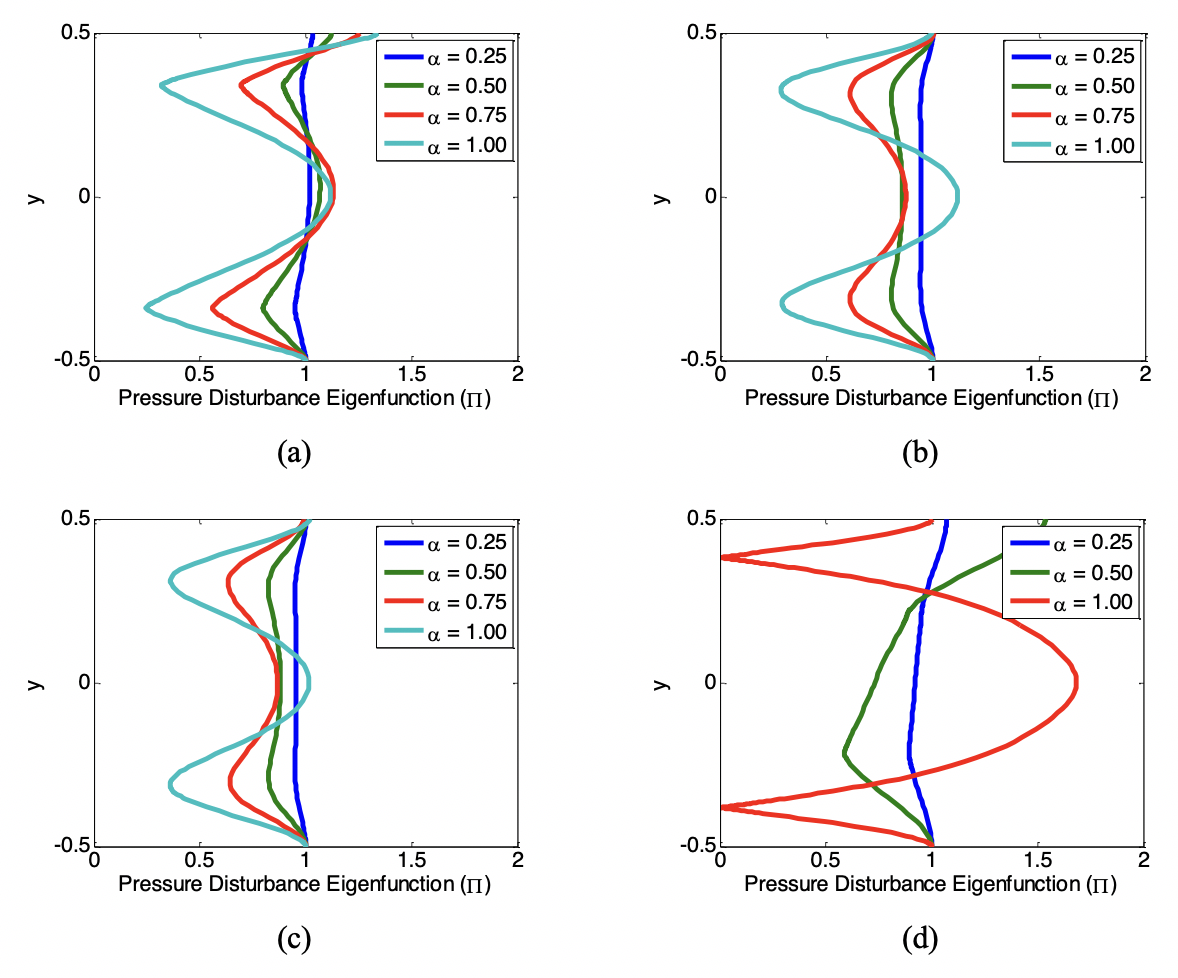}
    \caption{The variation of pressure disturbance eigenfunction with respect to wavenumber and space has been shown for (a) Station 1, (b) Station 2, (c) Station 3, and (d) Station 4.  \label{figure_5}}
\end{figure}

We see from figure \ref{figure_5} that the pressure disturbance eigenfunction shows some particularly interesting trends. Firstly, the qualitative nature of the eigenfunction at all wavenumbers and at all spatial locations is similar. There exist two locations along the transverse co-ordinate where the disturbance value reaches its maximum. There is some asymmetry in the locations as the LES data itself is not completely symmetric about the centerline. These maxima may be attributed to the presence of critical layers. If the quantity $U - c$ vanishes somewhere in the flow domain, Rayleigh's equation (\ref{equation_17}), or equivalently (\ref{equation_18}), becomes singular in that the term multiplying the highest derivative vanishes, unless $U''(y_c) = 0$. This occurs at a point $y_c$, say, when the phase speed $c$ is real. Thus, $y_c$ is a singular point and defines the location known as the critical layer. Note that when the phase speed is complex the expression $U - c$ is no longer zero, and Rayleigh's equation is no longer singular. In our case, since we have a purely complex phase speed at these locations, the eigenfunction shoots up but does not blow up. Secondly, the absolute value of the disturbance is seen to have very similar values except at station 4. This sudden change of behavior may be attributed to the value of the phase speed at the wavenumbers chosen. The wavenumbers chosen are those where the phase speed and base flow velocity are very nearly equal throughout the domain. Note that the velocity is extremely small at station 4 as compared to all the other stations. This causes the quantity $U - c$ to be very near to zero in most parts of the domain. Thirdly, the disturbances increase with increasing wavenumber. It is to be noted that the growth rate describes the rate of growth of the pressure disturbance eigenfunction and not its amplitude. Thus, there is no correlation between a maximum amplitude pressure eigenfunction and the fastest growing eigenfunction. The existence of critical layers is proven in the next section. 

We have not provided here the variation of the stream-wise and transverse velocity fluctuations. The trend of the eigenfunctions is the same as expected. There is indeed a sudden increase of the eigenfunctions due to the critical layers.

\subsection{Critical Layers}
Critical layers arise as a singularity of the linearized Euler equations when the phase speed of the disturbance is equal to the mean flow velocity. The sudden increase in the disturbance eigenfunctions at particular locations on the transverse co-ordinate are typical signatures of presence of critical layers. As explained previously, critical layers occur if the quantity $U - c$ vanishes somewhere in the flow domain. Since the phase speed is complex, the expression $U - c$ is no longer zero but a very small number and thus the eigenfunctions see such a sudden increase. 

\begin{figure}
    \centering
    \includegraphics[width=\textwidth]{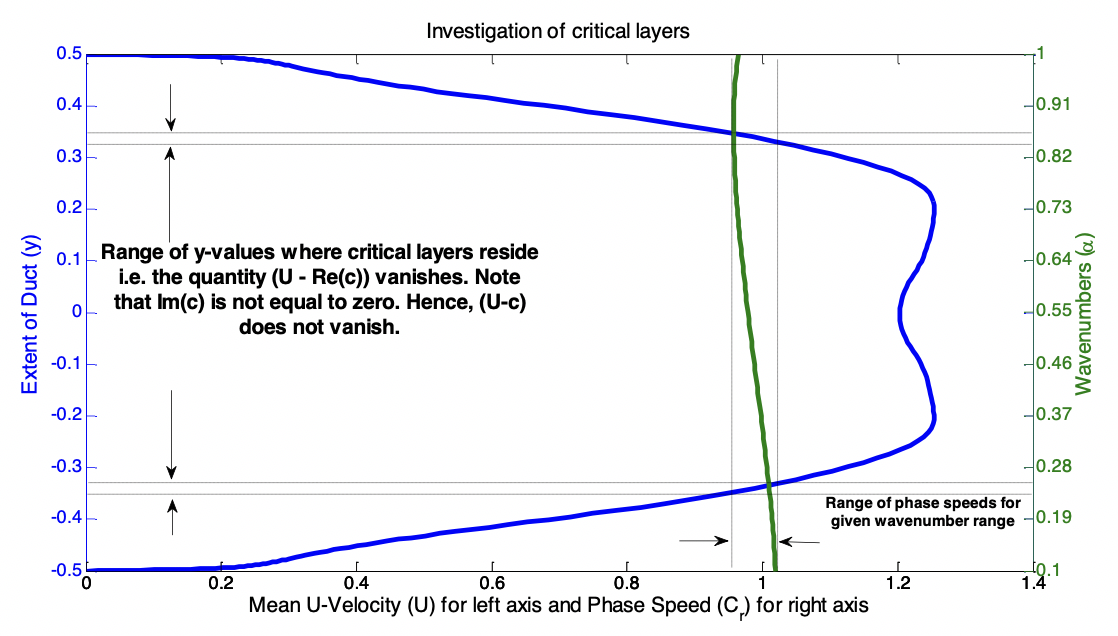}
    \caption{Superposition of the variation of phase speed with wavenumber and the velocity profile at station 2. \label{figure_6}}
\end{figure}

Figure \ref{figure_6} shows the superposition plot of the phase speed ($c_r$) with varying wavenumber along with the actual velocity profile at station 2 which has been obtained from the LES calculations. It can be clearly seen that the phase speed is bounded by the extrema of the velocity profile. This implies that critical layers will exist at all wavenumbers.

\subsection{Instability as a consequence of Rayleigh's Inflection Point Theorem}
Suppose that $u_B$ and $\frac{du_B}{dx}$ are continuous in $y_1 < y < y_2$. Then, Rayleigh's inflection point theorem states that a necessary (though not sufficient) condition for inviscid instability is that the base state possesses an inflection point $\frac{d^2u_B}{dy^2}=0$ somewhere in the domain $y_1 < y < y_2$. If a base state lacks an inflection point, we can conclude it to be stable, for inviscid fluids. The generalized inflection point is a point where the quantity $\frac{d}{dy}\left(\rho\frac{du_B}{dy}\right) = 0$. Figure 7 shows the generalized inflection points of the flow profile at station 2.

\begin{figure}
    \centering
    \includegraphics[width=\textwidth]{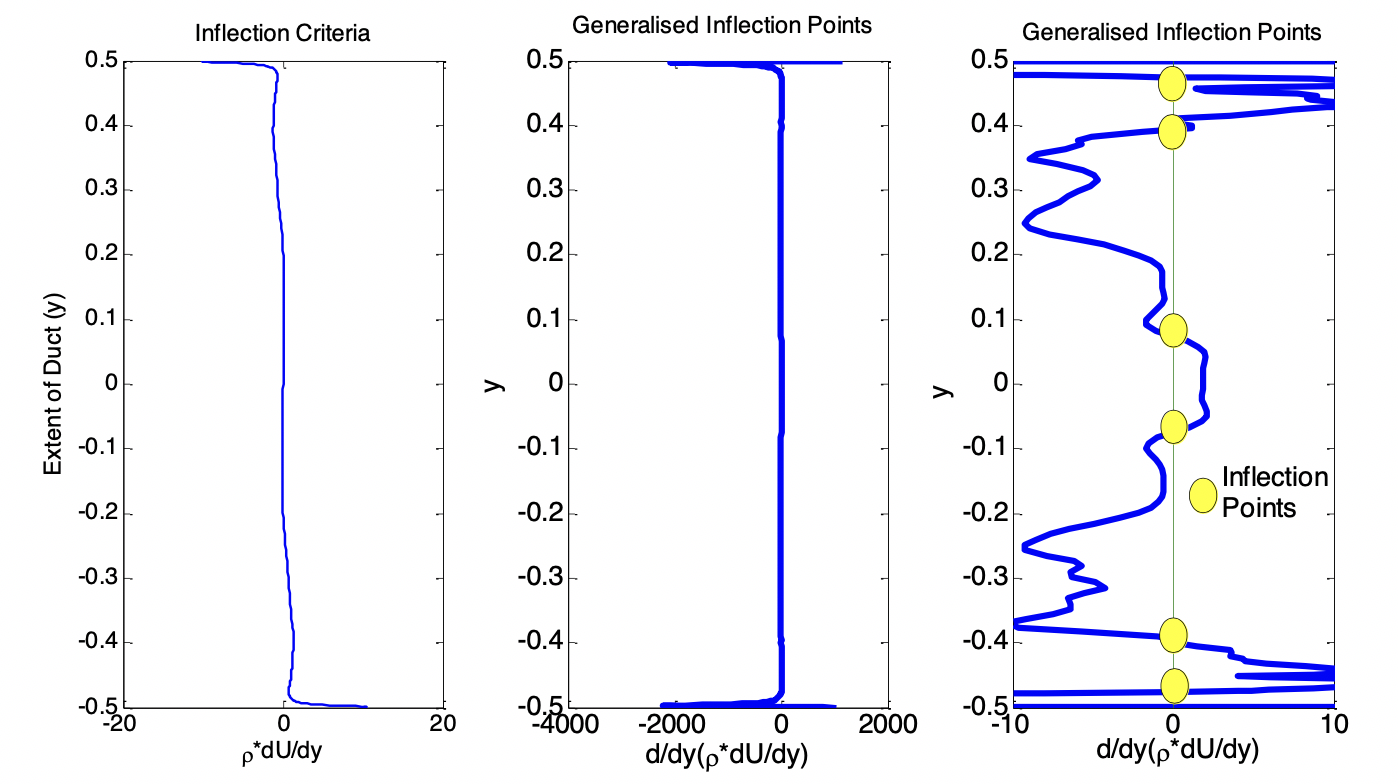}
    \caption{A plot of the generalized inflection points of the velocity profile at station 2. The first figure on the left shows a plot of the quantity $\rho\frac{du_B}{dy}$ with $y$. The figure in the middle shows a plot of the variation of the quantity $\frac{d}{dy}\left(\rho\frac{du_B}{dy}\right)$ with $y$. The last figure shows a zoomed view of the variation of $\frac{d}{dy}\left(\rho\frac{du_B}{dy}\right)$ with $y$ near zero, so that the inflection points are clearly visible.  \label{figure_7}}
\end{figure}

\section{Discussion of Results}

\subsection{Validity of the parallel flow assumption}
\begin{figure}
    \centering
    \includegraphics[width=0.9\textwidth]{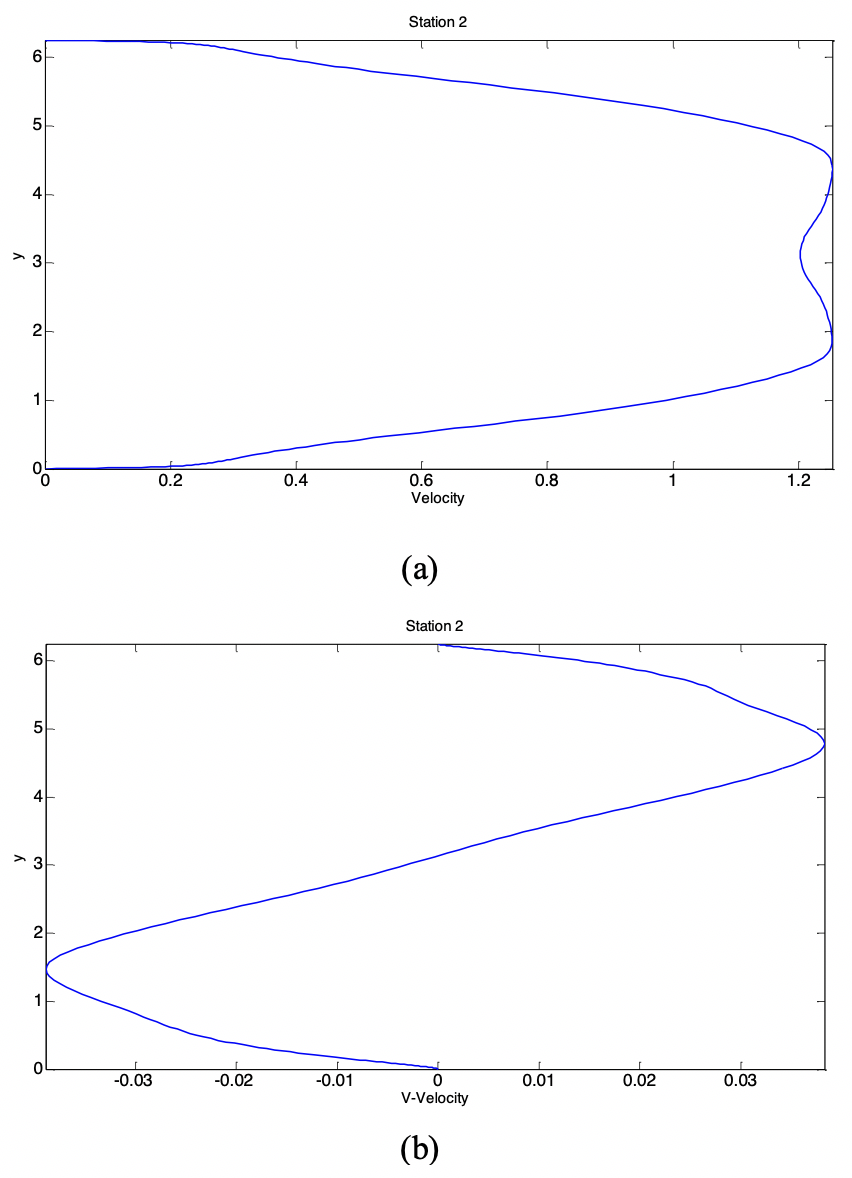}
    \caption{Plot of (a) $u$ and (b) $v$ variation in $y$ at Station 2. Velocities have been normalized by $c$.  \label{figure_8}}
\end{figure}

We have assumed in the derivation of the linearized perturbation equations that the base flow is a parallel flow. The base flow in our case is matched to the velocity profiles that are obtained at the selected stations from the LES calculations. To that end, let us plot the $u$ and $v$ velocities at station 2 and check their relative magnitudes. Note that the reference parameter for both $u$ and $v$ is the speed of sound $c = \sqrt{\gamma R^* T^*}$ where $T^*$ is the reference temperature, $R^*$ is the gas constant for air and $\gamma$ is the ratio of the specific heats of air. The plots of $u$ and $v$ are shown for station 2 ($x = 33.1482 \delta_1$) in figure \ref{figure_8}.

From the figures above, we can see that $\max \{u\} = 1.2538$ and $\max \{v\} = 0.0383$. Thus, the ratio of $v$ to $u$ is
\begin{equation*}
    \frac{v_{\text{max}}}{u_{\text{max}}} = \frac{0.0383}{1.2538} = 0.0305 = 3.05\%
\end{equation*}
This ratio keeps on increasing to about 20\% as we go further downstream of the shock train. This is a serious critique of the parallel flow assumption, as was already observed earlier.

\subsection{Dimensional Length of the Most Unstable Disturbance}
We wish to find the dimensional length of the most unstable disturbance. In the following analysis, starred `*' quantities denote dimensional quantities and non-starred quantities are non-dimensional quantities.
\begin{enumerate}
    \item Reference length for governing equations is $L$. Hence, let $x = \frac{x^*}{L}$ and $y = \frac{y^*}{L}$
    \item The LES of a normal shock train in a constant-area isolator model is carried out with $M_\infty = 1.61$, $Re_\theta = 1660$, and $Re_\delta = 16200$ \cite{Morgan2012,Duraisamy12}.
    \item The reference length $L$ for our stability analysis is the width of the duct $= 6.25 \delta_1$ where $\delta_1$ is the boundary layer thickness at the span-wise position in the duct where the boundary layer thickness $\delta = 5.4$ mm (thus, $\delta_1 = 5.4$ mm) as shown in figure 9. This reference position $x_1$ matches with the position $x_{\text{in}} = 264.8$ mm in the Caroll \cite{Carroll88} experiment. In the experimental 750 mm long, nominally rectangular test section, LDV measurements were taken beginning at $x_{\text{in}} = 264.8$ mm (at approximately the location of the initial normal shock) and extending downstream at variable intervals over 400 mm.
    \item Non-dimensional wavenumber $= \alpha = \frac{2\pi}{\lambda}$ and $\lambda = \frac{\lambda^*}{L}$. Now, maximum growth rate occurs at $\alpha = 0.65$. This means that the corresponding $\lambda = 2\pi/0.65 = 9.6664$. The wavelength of the maximum disturbance is then: $9.6664*6.25 \delta_1 = 60.4150 \delta_1 = 60.4150$ x $5.4 = 326.24$ mm $\approx$ Computational Domain Length $> 4 \delta_1 = 21.6$ mm $=$ Distance between first and second shock. 
\end{enumerate}
Thus, we can conclude that the most linearly unstable disturbance is almost 0.75 times the length of the computational domain (shock train region). A physical disturbance is anticipated to be of the size of the distance between two consecutive shocks. However, the disturbance that our analysis yields is in some sense `unphysical' as it does not conform to the physical length scales of the problem.

\begin{figure}
    \centering
    \includegraphics[width=\textwidth]{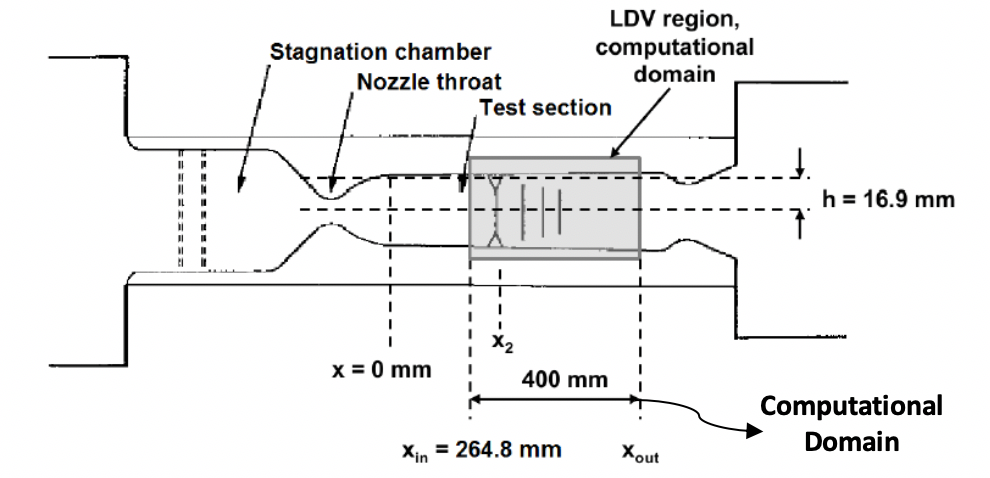}
    \caption{A schematic of the experimental setup used by Carroll and Dutton \cite{Carroll88}. The $400$ mm section denotes the computational domain over which the LES calculation has been performed.  \label{figure_9}}
\end{figure}

\section{Conclusions}
Temporal instability is observed at all the stations beyond station 1. However, all disturbances are seen to propagate downstream. There exist inflection points in the velocity profile which give rise to the instabilities as suggested by Rayleigh's theorem. The growth rate of these instabilities is significantly large. 

There exist two narrow regions where the flow velocity equals the phase speed of the disturbance. These are called critical layers. These critical layers are seen to exist at all stations for all wavenumbers (non-dimensional) in the interval 0.1 to 1 that have been analyzed.
 
The ratio of the transverse velocity to the stream-wise velocity is about 3\% at the first station and increases downstream. This puts to test the parallel flow assumption in the inviscid stability calculations. The wavelength of the fastest growing disturbance is calculated to be about 326.24 mm whereas the computational domain itself is about 400 mm. Thus, we must further refine our calculations using techniques such as dynamic mode decomposition, an algorithmically efficient version of global stability analysis. This is the current avenue for future research on this subject.

\section{Acknowledgements}
The author of this technical report was advised by Prof. Sanjiva K. Lele to conduct this research between October 2012 and April 2013 at Stanford University. The author was funded by the School of Engineering Fellowship from Stanford University for this work. The author would like to sincerely thank the Predictive Science Academic Alliance Program (PSAAP II) at Stanford University for funding this work and subsequent work undertaken as part of the program. Computing time was provided by the High Performance Computing Center (HPCC) and the Center for Turbulence Research (CTR) at Stanford University, Stanford, CA on the Certainty cluster.  The author would also like to thank Dr. Carlo Scalo of the Center for Turbulence Research (CTR) at Stanford for many helpful discussions.

\appendix

\bibliographystyle{unsrt}  
\bibliography{References}

\end{document}